# Synthesis of Ultra-Incompressible Carbon Nitrides Featuring Three-Dimensional Frameworks of CN$_4$ Tetrahedra Recoverable at Ambient Conditions

Dominique Laniel,[1,2,*] Florian Trybel,[3] Andrey Aslandukov,[2,4] Saiana Khandarkhaeva,[2] Timofey Fedotenko,[5] Yuqing Yin,[2,6] Ferenc Tasnádi,[3] Alena V. Ponomareva,[7] Gunnar Weck,[8] Fariia Iasmin Akbar,[4] Bjoern Winkler,[9] Adrien Néri,[4] Stella Chariton,[10] Carlotta Giacobbe,[11] Jonathan Wright,[11] Gaston Garbarino,[11] Björn Wehinger,[11] Anna Pakhomova,[11] Mohamed Mezouar,[11] Vitali Prakapenka,[10] Victor Milman,[12] Wolfgang Schnick,[13] Igor A. Abrikosov,[3] Leonid Dubrovinsky,[4] Natalia Dubrovinskaia[2,3]

**Affiliations:**

[1]Centre for Science at Extreme Conditions and School of Physics and Astronomy, University of Edinburgh, EH9 3FD Edinburgh, United Kingdom

[2]Material Physics and Technology at Extreme Conditions, Laboratory of Crystallography, University of Bayreuth, 95440 Bayreuth, Germany

[3]Department of Physics, Chemistry and Biology (IFM), Linköping University, SE-581 83, Linköping, Sweden

[4]Bayerisches Geoinstitut, University of Bayreuth, 95440 Bayreuth, Germany

[5]Photon Science, Deutsches Elektronen-Synchrotron, Notkestrasse 85, 22607 Hamburg, Germany

[6]State Key Laboratory of Crystal Materials, Shandong University, Jinan 250100, China

[7]Materials Modeling and Development Laboratory, NUST "MISIS", 119049 Moscow, Russia

[8]CEA, DAM, DIF, F-91297 Arpajon, France

[9]Institut für Geowissenschaften, Abteilung Kristallographie, Johann Wolfgang Goethe-Universität Frankfurt, Altenhöferallee 1, D-60438 Frankfurt am Main, Germany.

[10]Center for Advanced Radiation Sources, University of Chicago, Chicago, Illinois 60637, United States

[11]The European Synchrotron Radiation Facility, 38043 Grenoble Cedex 9, France

[12]Dassault Systèmes BIOVIA, CB4 0WN Cambridge, United Kingdom

[13]Department of Chemistry, University of Munich (LMU), Butenandtstrasse 5-13, 81377 Munich, Germany

*Correspondence to: dominique.laniel@ed.ac.uk

**Abstract**

More than thirty years ago, carbon nitrides featuring 3D frameworks of tetrahedral CN$_4$ units were identified as one of the great aspirations of materials science, expected to have a hardness greater than or comparable to diamond[1]. Since then, no unambiguous experimental evidence of their existence has been delivered. Here, we report the high-pressure high-temperature synthesis of the long-sought-after covalent carbon nitrides, $tI$14-C$_3$N$_4$, $hP$126-C$_3$N$_4$, and $tI$24-CN$_2$, in laser-heated diamond anvil cells. Their structures were solved and refined using synchrotron single-crystal X-ray diffraction. In these solids, carbon atoms—all $sp^3$-hybridized—and nitrogen atoms are fully saturated, forming four and three covalent bonds, respectively, leading to three-dimensional arrangements of corner-sharing $CN_4$ tetrahedra. These carbon nitrides are ultra-incompressible, with $hP126$-$C_3N_4$ and $tI24$-$CN_2$ even rivalling diamond's incompressibility, and superhard. These novel compounds are recoverable to ambient conditions in crystalline form and chemically stable in air. Being wide-band gap semiconductors with intriguing features in their electronic structure, they are expected to exhibit multiple exceptional functionalities besides their mechanical properties, opening new perspectives for materials science.

**Introduction**

In 1989 it was predicted that "hypothetical covalent solids formed between carbon and nitrogen are good candidates for extreme hardness"[1]. A $C_3N_4$ solid isostructural to β-$Si_3N_4$, chosen by Liu and Cohen[1] as a prototype system, was calculated to have a bulk modulus of 427 GPa[1], rivalling that of diamond (446 GPa)[2]. These predictions sparked enormous interest from both experimental and theoretical communities[3–25] that has led for now to more than 6000 scientific articles[26]. In particular, it was suggested that the initially considered β-$Si_3N_4$-type $C_3N_4$[1] is thermodynamically unfavourable compared to other C-N compounds[15,23]. Three possible stoichiometries, CN, $C_3N_4$, and $CN_2$, were proposed for stable carbon nitrides, all featuring 3D framework structures in which N atoms and $sp^3$-hybridized C atoms are three- and fourfold-coordinated, respectively, and thus clearly distinct from graphitic polymeric C-N compounds.[26] These are all expected to be superhard and possess excellent thermal conductivity, wide bandgaps, and exotic electronic properties[15]—confirming the exceptional potential of this class of materials. Moreover, considering the great perspectives for diamond-based electronics[27,28], novel semiconducting materials based on easily accessible and environmentally-friendly elements (*e.g.* C and N) with an exceptional combination of properties including variable band gaps, are highly desirable.

Intense efforts were devoted to synthesizing these promising carbon nitrides. Various experimental techniques were employed, in particular, chemical and physical vapor depositions[29], solvothermal methods[10], and static and dynamic high-pressure methods[4,8,9,11,14,18,19,30,31]. There is, however, a single report on the formation of a fully saturated $sp^3$-hybridized carbon nitride, namely the CN compound, obtained in a diamond anvil cell laser-heated to 7000 K above 55 GPa[4]. Still, the structure was proposed based on the match of the powder diffraction peaks of the observed CN with those of the theoretically predicted β-InS type CN compound[4]. Rietveld refinement was not possible because of "preferred orientation effects and strongly anisotropic peak broadening"[4]. Moreover, this CN compound was reported non-recoverable to ambient conditions, found unstable below 15 GPa[4].

Here, we report the synthesis of three covalent carbon nitrides with the chemical compositions $C_3N_4$ and $CN_2$, that have never been observed experimentally as yet. Their synthesis was realised at 124 and 134 GPa and at temperatures near 2500 K in laser-heated diamond anvil cells. The crystal structures of these compounds, $tI14$-$C_3N_4$, $hP126$-$C_3N_4$, and $tI24$-$CN_2$, were solved and refined and their ultra-incompressibility was revealed on the basis of synchrotron single-crystal X-ray diffraction (SC-XRD) data. All of the compounds feature $sp^3$-hybridized carbon; the corner-sharing $CN_4$ tetrahedra form 3D frameworks. The previously observed β-InS type CN compound ($oP8$-CN) was synthesised at 72 GPa and 2500 K and its structure is now solved and refined based on SC-XRD. All four carbon nitrides are found to be stable in air at ambient conditions. Density functional theory (DFT)-based calculations reported in this work predict these carbon nitrides to be superhard and possess functionally-promising wide band gaps.

**Results and Discussion**

Three screw-type BX90 diamond anvil cells (DACs)[32] were prepared to investigate the carbon-nitrogen system in a pressure range of about 70 to 137 GPa. Two DACs were loaded with tetracyanoethylene (TCNE, $C_6N_4$, Alfa Aesar, 98% purity) and gas-filled with nitrogen (sample #1 and sample #2). The third DAC was also gas-filled with nitrogen and a small piece of black phosphorus put in this cell served as a light absorber for laser heating (sample #3). In each DAC, the sample was first pressurized to the desired pressure and then laser heated. Figure S1 shows microphotographs of the three samples before and after laser-heating. In all DACs chemical reactions were observed, which led to the formation of new C-N compounds, whose structures, described in detail below, were solved and refined from SC-XRD.

The laser-heating at 2700 K of the sample #1 (TCNE and $N_2$) at 124 GPa resulted in the formation of two novel carbon nitrides with chemical composition $C_3N_4$, each with a different structure. Using the Pearson symbols notation, these compounds are labelled as $tI$14-$C_3N_4$ and $hP$126-$C_3N_4$. The full crystallographic data is provided in Table S1 and Table S2.

The structure of $tI$14-$C_3N_4$ has a tetragonal unit cell (space group $I$-42$m$, #121) with parameters $a$ = 3.2285(6) Å and $c$ = 6.4526(16) Å ($V$ = 67.26(3) Å$^3$) at 124 GPa. It is composed of three crystallographically distinct atoms, C1, C2, and N1, on the 2$a$, 4$d$, and 8$i$ Wyckoff sites, respectively. Both carbon atoms are fourfold coordinated by nitrogen atoms, forming the long-sought-after $CN_4$ tetrahedron, and the N1 atom is threefold coordinated by carbon atoms (Figure 1a). The C1-N1 and C2-N1 bonds are very similar in length, being respectively of 1.381(2) Å and 1.399(2) Å. The average N-C-N bond angle is 109.5(1)°, perfectly matching the value expected of a regular tetrahedron, highlighting the $sp^3$-hybridization of the carbon atoms. The average C-N-C bond angle being of 110.0(1)° also suggests $sp^3$-hybridization of the nitrogen atoms. Corner-sharing $CN_4$ tetrahedra form a three-dimensional network, which is easy to visualise (see Figure 1b,c), if the structure of $tI$14-$C_3N_4$ is considered as a defect variant of sphalerite (ZnS) structure type—which is that of cubic boron nitride as well. It consists of layers of tetrahedra stacked in an ABC sequence (Figure 1b) in the [1$\bar{1}$2] direction, but instead of complete layers as in ZnS, the layers in $tI$14-$C_3N_4$ are incomplete (Figure 1c): each fourth tetrahedron in a row is missing. Thus, $tI$14-$C_3N_4$ is isostructural to $\beta$-$Cu_2HgI_4$[33].

Some single-crystal datasets of $tI$14-$C_3N_4$ revealed supercell reflections. The largest superstructure that could be identified has an orthorhombic unit cell (space group $Pma$2, #28) with lattice parameters $a$ = 13.6827(14) Å, $b$ = 13.6728(11) Å and $c$ = 19.346(4) Å ($V$ = 3619.2(9) Å$^3$) at 130 GPa, as seen in Figures S2 and S3. An experimental crystal model for this supercell could not be obtained due to an insufficient number of reflections.

The structure of $hP$126-$C_3N_4$ has a large hexagonal unit cell (space group $P6_3/m$, #176) with lattice parameters of $a$ = 17.690(8) Å and $c$ = 2.2645(11) Å ($V$ = 613.7(5) Å$^3$) at 124 GPa (Figure 1c). It is composed of 21 crystallographically distinct atoms, nine C and twelve N, all on 6$h$ Wyckoff sites. Like in $tI$14-$C_3N_4$, carbon and nitrogen atoms in $hP$126-$C_3N_4$ are fully saturated and tetrahedrally and triply coordinated by N and C atoms, respectively. A polyhedral model of $hP$126-$C_3N_4$ viewed along the $c$ direction is shown in Figure 1d. The corner-sharing $CN_4$ tetrahedra form a rather complex arrangement. Some features of this structure are reminiscent of $\beta$-$Si_3N_4$[34], namely the six-membered rings of tetrahedra (*sechser*-rings, according to Liebau[35] nomenclature), which form large empty channels in the $c$-direction centered at the origin of the unit cell. However, unlike $\beta$-$Si_3N_4$, which is exclusively composed of interlinked *sechser*-rings, $hP$126-$C_3N_4$ is made up of interlinked three- (*dreier*-rings), four- (*vierer*-rings), and five-membered rings (*fünfer*-rings).

The average C-N bond length was determined to be 1.376(5) Å, which is, within uncertainty, identical to the C1-N bond in $tI$14-C$_3$N$_4$. The average N-C-N bond angle is 109.43(7)°—analogous to that in $tI$14-C$_3$N$_4$, pointing to the $sp^3$-hybridization of carbon. The C-N-C bond angle varies between 108.31(2) and 124.5(4)° (with one outlier of 132.6(3)°), the average bond angle being 117.5(1)°, which suggests predominantly $sp^2$-hybridized nitrogen atoms with a minor fraction being $sp^3$-hybridized, in contrast to fully $sp^3$-hybridized N in $tI$14-C$_3$N$_4$. Interestingly, the three carbon and three nitrogen atoms lay in the same $ab$ plane forming flat C$_3$N$_3$ rings (highlighted by red circles in Figure 1 d), reminiscent of triazine rings in heptazine and its derivatives (namely melem, C$_6$N$_7$(NH$_2$)$_3$)[36], albeit solely composed of single bonds. The $hP$126-C$_3$N$_4$ phase, having a larger volume per atom and higher enthalpy than $tI$14-C$_3$N$_4$ at 124 GPa, is deduced to be formed at higher temperatures while $tI$14-C$_3$N$_4$ is produced at lower temperatures (see Supplementary Materials).

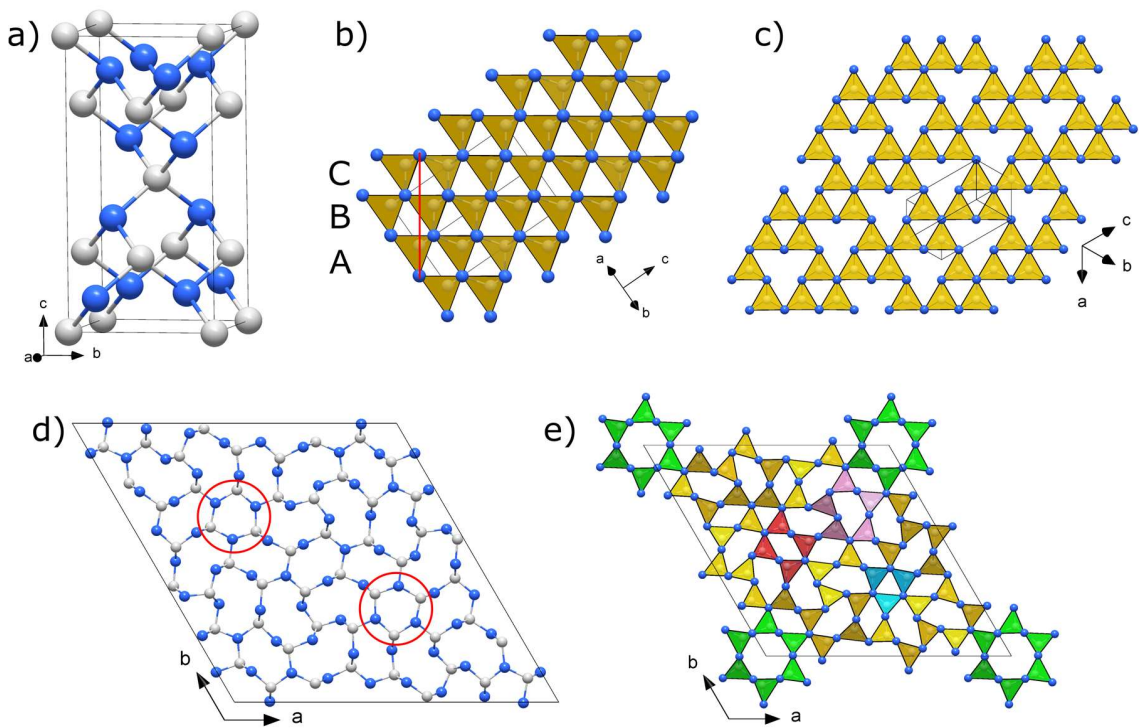

**Figure 1: Crystal structures of $tI$14-C$_3$N$_4$ and $hP$126-C$_3$N$_4$ at 124 GPa.** a) Unit cell of $tI$14-C$_3$N$_4$. b) View of the polyhedral model of the $tI$14-C$_3$N$_4$ structure along the [110] direction; the structure built of corner-sharing CN$_4$ tetrahedra can be interpreted as an ABC stacking of layers (oriented perpendicular to the page) in the [1$\bar{1}$2] direction (marked by a red line); c) a single layer viewed along the [1$\bar{1}$2] direction. d) A projection of the unit cell of $hP$126-C$_3$N$_4$ on the $ab$ plane; flat C$_3$N$_3$ rings, are highlighted by red circles. e) A polyhedral model of $hP$126-C$_3$N$_4$ viewed along the $c$ direction. The green, pink, red, and blue sets of tetrahedra highlight the six-, five-, four- and three-membered groups of CN$_4$ tetrahedra. Grey and blue spheres represent carbon and nitrogen atoms, respectively.

The other TCNE and N$_2$ sample (sample #2) was heated to about 2600 K at 72 GPa. At this pressure, the diffraction pattern of the resulting C-N compound was indexed in an orthorhombic unit cell (space group $Pnnm$, #58) with $a$ = 4.892(4) Å, $b$ = 3.7841(12) Å and $c$ = 2.2880(9) Å ($V$ = 42.36(4) Å$^3$). The quality of single-crystal XRD data was sufficient for the structure solution and refinement (Figure 2 a,b,

Table S3). The compound was found to have the CN stoichiometry and the β-InS structure type[37] (*oP*8-CN). Such a compound was previously reported to be produced by Stavrou *et al*.[4], formed by laser-heating of graphite in molecular nitrogen at pressures above 55 GPa and 7000 K[4], but its structure could not be refined based on that time only available powder diffraction data. The structure of *oP*8-CN is described in the Supplementary Materials.

One more hitherto unknown C-N phase, namely *tI*24-CN$_2$, was synthesised in the DAC containing phosphorus and nitrogen (sample #3) after its laser-heating to ~2500 K at 134 GPa due to a reaction of nitrogen with carbon from the diamond anvils. A number of other phases were also observed in this DAC, namely the well-known cg-N and bp-N polymeric nitrogen allotropes[38,39], the incommensurate phase IV of phosphorus[40], and phosphorus nitride PN$_2$[41].

The structure of the new *tI*24-CN$_2$ carbon nitride has a tetragonal unit cell ($I\bar{4}2d$ space group, #122) with the lattice parameters $a$ = 5.9864(13) Å and $c$ = 3.230(2) Å ($V$ = 115.75(8) Å$^3$) at 134 GPa. The high-quality single-crystal XRD data of this phase enabled a complete structure solution and refinement (Table S4). *tI*24-CN$_2$ is found to be composed of two crystallographically unique atoms: C1 (8*d*) and N1 (16*e*). Carbon and nitrogen are, respectively, fourfold and threefold coordinated, but not all covalent bonds are heteroatomic. Indeed, while carbon is solely making C-N bonds—forming a CN$_4$ tetrahedron—nitrogen has also an N-N bond (Figure 2c,d). Thus, the structure of *tI*24-CN$_2$ can be presented as a framework of CN$_4$ tetrahedra connected through N$_2$ dimers. When looking along the *c*-axis (Figure 2c), the crystal structure can be described as *vierer*-rings interconnected through their corners and N$_2$ dimers, and repeating in the [001] direction, forming square-shaped channels along this direction. The four C-N bonds of the CN$_4$ tetrahedra are identical within their uncertainty, with two of 1.379(7) Å and two others of 1.389(7) Å; bond lengths very similar to those of *oP*8-CN, *tI*14-C$_3$N$_4$ and *hP*126-C$_3$N$_4$. The N-C-N bond angles of the CN$_4$ tetrahedra vary between 104.0(3)° and 116.7(6)°, with the average value being 109.5(3)°—in line with *sp*$^3$-hybridization within regular CN$_4$ tetrahedra. The N-N bond length is 1.272(11) Å; slightly shorter than what is expected from a single bond in 3D nitrogen polymeric solids, found to have a length of 1.3410(15) Å in cg-N (122 GPa)[42] and of 1.338(6) Å and 1.435(7) Å in bp-N (140 GPa)[39]. Still at 134 GPa, *tI*24-CN$_2$ has an average angle formed by the C/N-N-C bonds of 117.3(4)°, suggesting predominantly *sp*$^2$-hybridization of nitrogen.

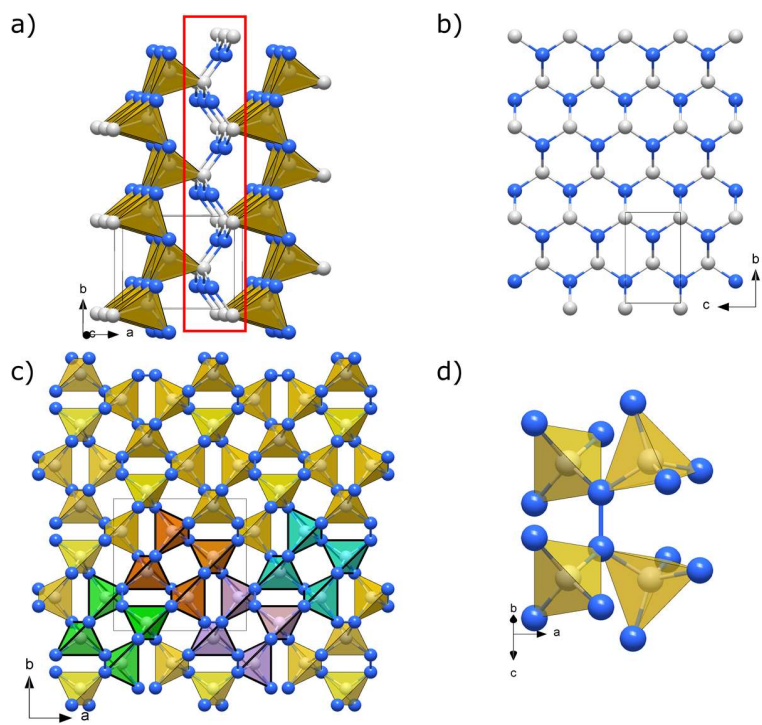

**Figure 2: Crystal structures of *oP*8-CN and *tI*24-CN$_2$ at 72 and 134 GPa, respectively.** a) A polyhedral model of the *oP*8-CN structure built of C(CN$_3$) tetrahedra. The tetrahedra sharing nitrogen vertices form corrugated layers laying in the *bc* plane, which are connected through carbon apexes of the tetrahedra by triply coordinated nitrogen atoms, as highlighted by the red rectangle. b) A corrugated honeycomb-like net of the 1:1 C:N composition formed by the atoms connecting the layers. c) A polyhedral model of the *tI*24-CN$_2$ structure (viewed along the *c* direction) built of corner-sharing CN$_4$ tetrahedra linked with each other through N$_2$ dimers. To emphasize that the crystal structure can be understood as repeating units composed of four CN$_4$ tetrahedra, four of these units are drawn in different colors (green, orange, purple, and teal). d) An example of an N$_2$ dimer (oriented vertically in the figure) connecting two pairs of corner-sharing CN$_4$ tetrahedra. Grey and blue spheres represent carbon and nitrogen atoms, respectively.

Whereas three carbon nitrides synthesised here have been theoretically predicted in a number of works (*tI*14-C$_3$N$_4$[15,18], *oP*8-CN[15,43], and *tI*24-CN$_2$[15,24]), the phase with the most complex structure and a large unit cell, *hP*126-C$_3$N$_4$, has never been proposed. Our DFT calculations show that the relaxed theoretical structural models perfectly reproduce the experimental models (Tables S1-S4) and that the *hP*126-C$_3$N$_4$ solid is dynamically stable (Figure S4). Previous calculations have demonstrated the dynamical stability of the *tI*14-C$_3$N$_4$, *oP*8-CN, and *tI*24-CN$_2$ compounds[15,24,43]. Our calculations also suggest all four carbon nitrides to be dynamically stable at ambient conditions (see Figure S4 and S5), providing a strong incentive to investigate these compounds under decompression.

After synthesis of the carbon nitrides, the pressure in all cells was gradually decreased and the behavior of the four solids was monitored using SC-XRD, allowing to obtain their pressure-volume relationship on the decompression (Figure 3, Table S5). Ambient conditions were reached by opening the three DACs, exposing the samples to air. High-quality SC-XRD data could still be collected, which revealed that all of the nitrides, *i.e.* *tI*14-C$_3$N$_4$, *hP*126-C$_3$N$_4$, *oP*8-CN, and *tI*24-CN$_2$, not only sustained at ambient conditions and air, but preserved their crystallinity and high-pressure crystal structures (see the crystallographic data at 1 bar in Tables S6-S9). This case is unique for materials synthesised at about 100

GPa and above—to the best of our knowledge similar behavior has not been reported before—and opens up very important perspectives not only for C-N compounds but for high-pressure materials science in general.

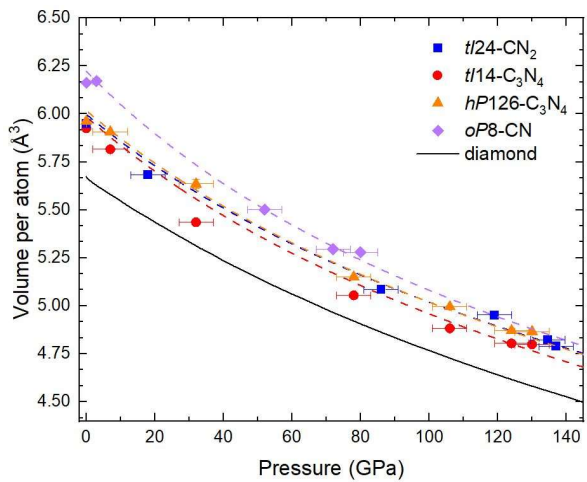

**Figure 3: Experimental and calculated pressure dependence of the unit cell volume per atom for the *tI*14-C$_3$N$_4$, *hP*126-C$_3$N$_4$, *oP*8-CN, and *tI*24-CN$_2$ solids found in this work and diamond, according to ref. 2.** The full symbols represent experimental data points obtained from SC-XRD data; the dashed lines of the corresponding colours are fits of the DFT data (see Supplementary Material for details) with the third-order Birch-Murnaghan equation of state (Table S10). The black curve is the data for diamond[2]. An uncertainty of ±5 GPa is given for all pressures but ambient, while the uncertainty on volume is smaller than the size of the points.

Based on their crystal chemistry, all four carbon nitrides found in this work are expected to be very incompressible. Fitting the p-V data obtained in the whole pressure range studied, using the second-order Birch-Murnaghan equation of state[44] with V$_0$ fixed on the experimental values at 1 bar, yields the following bulk moduli: for *oP*8-CN, K$_0$ = 365(9) GPa (V$_0$ = 49.30 Å$^3$); for *tI*14-C$_3$N$_4$, K$_0$ = 383(9) GPa (V$_0$ = 82.95 Å$^3$); for *hP*126-C$_3$N$_4$, K$_0$ = 417(6) GPa (V$_0$ = 751.2 Å$^3$); for *tI*24-CN$_2$, K$_0$ = 419(8) GPa (V$_0$ = 142.79 Å$^3$). These values are in good agreement with those obtained from DFT calculations (see Table S10). All four carbon nitrides thereby classify as ultra-incompressible materials, with the bulk moduli of *tI*24-CN$_2$ and *hP*126-C$_3$N$_4$ being greater than that of cubic boron nitride c-BN (395(2) GPa)[45] and comparable to that of diamond (446 GPa)[2]. Considering the empirical correlation between incompressibility and hardness[1,46], the synthesised carbon-nitrogen compounds are expected to be superhard. Our hardness calculations (see Table S11) using different hardness models (microscopic[47] and macroscopic[48]) confirm that they all indeed belong to the class of superhard materials, with values comparable to or even exceeding the hardness of c-BN, and diamond—in accordance with the early predictions of Liu and Cohen[1].

The carbon nitrides synthesized in this work are expected to exhibit multiple exceptional functionalities besides their mechanical properties, with a potential to be ultimate engineering materials in the same category as diamond[27]. Considering their very high atomic densities (0.162-0.169 atoms/Å$^3$ vs 0.176 atoms/Å$^3$ for diamond), they are expected to be excellent thermal conductors[23,49,50]. They are also anticipated to be transparent over a wide range of wavelengths: visual observations of the *tI*24-CN$_2$ and *hP*126-C$_3$N$_4$ samples provide evidence for their transparency (Figure S1) and theoretical calculations of their bandgaps at ambient conditions (see Table S11) show that all four carbon nitrides are wide band gap semiconductors comparable to diamond. Indeed, the band gaps are of 5.48, 4.35, 4.43, 5.44 and 5.48 eV[51] for $oP8$-CN, $tI14$-$C_3N_4$, $hP126$-$C_3N_4$, $tI24$-$CN_2$ and diamond, respectively. Details of their electronic properties (Figure S6 to Figure S9) differ from one to another, which could make them adapted for specific applications not currently feasible for diamond-based electronics. For instance, while CN is an indirect band gap semiconductor like diamond, $CN_2$ could be a direct band gap material. Moreover, in the electronic structure of $tI14$-$C_3N_4$ and $hP126$-$C_3N_4$ one can clearly identify flat bands in the very vicinity of the band gap. In general, flat bands with energies that vary weakly with electron momentum favour strong correlations between the electrons, which in turn can yield exotic materials properties[52]. Variability of the compositions and crystal structures of the synthesized carbon nitrides leads to distinctive local chemical environments of C and N atoms, allowing one to expect different properties of native and external defects in these materials. This could potentially provide an answer to solving one of the major challenges of the diamond electronics in which the known dopants generate too deep energy levels. In addition, the variety of possible environments in these new wide band gap compounds is attractive for the search of new solid-state qubits, probably more stable than NV-centers in diamond. The non-centrosymmetric crystal structures of α-$C_3N_4$ and $CN_2$ could lead to important functionalities, such as piezoelectricity.

**Conclusions**

The high-pressure high-temperature synthesis and ambient conditions recovery of the ultra-incompressible $tI14$-$C_3N_4$, $hP126$-$C_3N_4$, $oP8$-CN, and $tI24$-$CN_2$ solids is the culmination of a three-decade-old quest for alternatives to diamond and c-BN. This study provides the *impetus* to further explore the rich chemistry of the carbon-nitrogen system and firmly establishes the exceptional mechanical and electronic properties of these materials. Our results also prove for the first time that materials synthesised at pressures over 100 GPa may be recovered at ambient conditions. Though immediate large-scale industrial production of carbon nitrides at megabar pressure range is not expected, given the current state of technology, the discovery that high-pressure C-N compounds are metastable at ambient conditions opens up non-trivial perspectives. For example, finding alternative synthesis pathways for these compounds—as done for high pressure $LiN_5$[53,54] or cg-N[38,55], for example—or the use of small amounts of crystals pre-synthesized in DACs as seeds for the growth of these phases in mild conditions in large-volume presses or by chemical vapor deposition (CVD), ion-beam deposition or reactive sputtering are viable possibilities.

**Methods**

*Experimental*

Three BX90 type diamond anvil cells[32] equipped with diamond anvil culets from 120 to 80 μm were prepared. Two of the cells (sample #1 and #2) were loaded with tetracyanoethylene (TCNE, $C_6N_4$, Alfa Aesar, 98% purity) and the other with a piece of black phosphorus (sample #3). All cells were also loaded with nitrogen gas at 1200 bars, serving as both a pressure transmitting media as well as a reactant. The pressure inside of the sample chamber was measured using the first order Raman mode of diamond[56] and verified using the calibrated diffraction lines of the rhenium gasket[57]. Laser heating was performed with double-sided Nd:YAG lasers (λ = 1064 nm) at our home laboratory at the Bayerisches Geoinstitut (BGI)[58] as well as at the GSECARS beamline of the APS using TCNE or phosphorus as laser absorbers. Temperatures were measured with an accuracy of ±200 K, using the thermoemission produced by the laser-heated samples[58].

Black phosphorus was synthesized out of red phosphorus using the 5000 tons uniaxial split sphere apparatus (Voggenreiter Zwick 5000[59]) at the Bayerisches Geoinstitut (BGI), under conditions of 2 GPa and 600°C maintained for 2 h. Lumps of red phosphorus (99.999%, Puratronics) were crushed in an agate mortar and subsequently loaded into a hexagonal boron nitride capsule to avoid any chemical reaction or oxidation during the synthesis. This capsule was then inserted into a 25/15 (octahedral edge length / anvil truncation edge length, in millimeter) BGI standard multi-anvil assembly equipped with a graphite heater. Temperature was monitored using a D-type thermocouple and kept constant during the time of the synthesis

The X-ray diffraction studies were done at the ID27 beamline ($\lambda$ = 0.3738 Å) and ID11 beamline ($\lambda$ = 0.2843 Å) of the Extreme Brilliant Source European Synchrotron Radiation Facility (EBS-ESRF) as well as at the GSECARS beamline of the APS ($\lambda$ = 0.2952 Å). In order to determine the position of the polycrystalline sample on which the single-crystal X-ray diffraction (SC-XRD) acquisition is obtained, a full X-ray diffraction mapping of the pressure chamber was achieved. The sample position displaying the most and the strongest single-crystal reflections belonging to the phase of interest was chosen for the collection of single-crystal data, collected in step-scans of 0.5° from −38° to +38°. The CrysAlis$^{Pro}$ software[60] was utilized for the single crystal data analysis. The analysis procedure includes the peak search, the removal of the diamond anvils' and other 'parasitic' signal contributions, finding reflections belonging to a unique single crystal, the unit cell determination, and the data integration. The DAFi program[61] was used for the automatic search of reflections' groups belonging to individual single-crystal domains. The crystal structures were then solved and refined using the OLEX2[62] and JANA2006 software[63]. OLEX2 was employed to obtain a preliminary structural model and JANA2006 to cull parasitic reflections (e.g. diamonds, other single-crystals) and obtain a final structural model. The SC-XRD data acquisition and analysis procedure was previously developed and described in detail in Ref. 64. Recently, this method was also successfully employed by other independent research groups[65–67].

Confocal Raman spectroscopy measurements were performed on three distinct setups. At the Commissariat à l'Énergie Atomique (CEA), an Alpha300M+ instrument (WITec) was employed with a continuous Ar–Kr laser using either the 488.0 or 647.1 nm lines with a focused laser spot of less than 1 μm. The Stokes Raman signal was collected in a back-scattering geometry by a CCD coupled to a 1800 l mm/1 grating, allowing a spectral resolution of approximately 1.5 cm$^{-1}$. Automated motorized sample positioning with piezo-driven scan stages of submicron accuracy allowed for precise Raman spectral imaging of the sample. At the Bayerisches Geo-Institut (BGI), a LabRam spectrometer equipped with a x50 Olympus long working distance objective was employed. For the sample excitation, a continuous He-Ne laser (632.8 nm) with a focused laser spot of about 2 μm in diameter was used. The Stokes Raman signal was collected in a backscattering geometry by a CCD coupled to a 1800 l/mm grating, allowing a spectral resolution of approximately 2 cm$^{-1}$. At the Advance Photon Source (APS), the GSECARS Raman system was utilized and an excitation wavelength of 532 nm selected. The details of this setup are described elsewhere[68]. Raman spectroscopy measurements were performed on all sample after their synthesis (see Figures S11-S14 of the Supplementary Materials).

*Ab initio calculations*

Kohn-Sham density functional theory based electronic structure calculations were performed with the QUANTUM ESPRESSO package[69–71] using the projector augmented wave method[72]. We used the generalized gradient approximation by Perdew-Burke-Ernzerhof (PBE) for exchange and correlation[73], with the corresponding potential files: for C and N the 1s electrons are treated as scalar-relativistic core states. We include van der Waals corrections following the approach by Grimme *et al.* as implemented in Quantum Espresso[74]. Convergence tests with a threshold of 1 meV per atom in energy and 1 meV/Å per atom for forces led to a Monkhorst-Pack[75] $k$-point grid of 12x8x16 for $oP8$-CN, 16x16x16 for $tI24$-CN$_2$, 16x16x8 for $tI14$-C$_3$N$_4$ and 2x2x16 for $hP126$-C$_3$N$_4$ with a cutoff for the wave-function expansion of 80 Ry for all phases.

We performed variable cell relaxations (lattice parameters and atomic positions) on all experimental structures to optimize the atomic coordinates and the cell vectors until the total forces were smaller than $10^{-4}$ eV/Å per atom and the deviation from the experimental pressure was below 0.1 GPa.

Equation of state (EOS) calculations were performed via variable-cell structural relaxations in 10 GPa steps up to 100 GPa for $oP8$-CN and 150 GPa for the other phases. We fitted a third-order Birch-Murnaghan EOS to the energy-volume points, calculated the P(V) and benchmarked versus the target pressure of the relaxations to ensure convergence.

Phonon dispersion relations (Figure S5) were calculated with Phonopy[76] in a 3x2x4, 3x3x3, 3x3x2 and 1x1x3 supercell for $oP8$-CN, $tI24$-CN$_2$, $tI14$-C$_3$N and $hP126$-C$_3$N$_4$ respectively. K-points have been adjusted according to the supercell size and reduced to 2x2x2 for $hP126$-C$_3$N$_4$.

As the size of the $hP126$-C$_3$N$_4$ unit cell combined with the strongly different lattice vector lengths ($a = b = 19.026$ Å, $c = 2.4180$ Å at ambient pressure) lead to limitations for the supercell size and geometry, therefore we decided to additionally benchmark the phonon calculations using the temperature-dependent effective potential method (TDEP[77,78]). Calculations were performed at the level of the harmonic approximation for the TDEP Hamiltonian including quantum and thermal effects at 300 K in a 1x1x3 supercell with respectively adjusted k-points at synthesis pressure (125 GPa) and ambient conditions (Figure S4). The calculations confirm the dynamical stability and are in very good agreement with the dispersion obtained with Phonopy at ambient conditions.

To estimate the hardness of the four carbon nitrides, we used two phenomenological models: a microscopic one developed by Lyakhov and Oganov al.[47] and a so called macroscopic model introduced by Chen et al.[48]. The former model is a microscopic hardness model because it is based on quantities such as the effective coordination number, the calculated bond lengths or the electronegativity values of each atoms, which are influenced by the local chemical environments. The chemical valences of C and N atoms and the covalent radii values that have been taken are as defined in the USPEX code[79]. To benchmark our implementation of the model we recalculated the hardness value of diamond (89.6 GPa)—as well as its bulk modulus (440 GPa)—and compared with the value in Ref. 48 (89.7 GPa). The second model is referred to as a macroscopic model since the hardness value is obtained from the material's calculated polycrystalline bulk (K) and shear (G) moduli (Table S12). To obtain these polycrystalline moduli of the four carbon nitrides, we calculated the anisotropic single-crystal elastic stiffness constants $C_{ij}$ (Table S13) and applied the Hill approximation. We used the energy-strain relationships and a finite difference method with (+/-) 1 and 2 % strain, where the total energies were calculated by Quantum Espresso with the numerical parameters given above. We furthermore calculated the directional dependence of the Young's modulus (Figure S10).

Electronic structure calculations (Figure S6) are performed with the generalized gradient approximation by Perdew-Burke-Ernzerhof (PBE) for exchange and correlation. Additional calculations for the band gaps are performed with the Heyd–Scuseria–Ernzerhof (HSE) hybrid functional[80] with the standard screening parameter and a $q$-grid of 4x4x4 for $oP8$-CN, 4x4x4 for $tI24$-CN$_2$, 4x4x2 for $tI14$-C$_3$N$_4$ and 2x2x2 for $hP126$-C$_3$N$_4$, with the k-points defining the band gaps with PBE being employed in the HSE sampling. The character of the gap does not change going from PBE to HSE, the size increases in general by ~1.5-2 eV (see Table S14). The crystal orbital bond index (COBI)[81], shown in Table S15, was calculated using the LOBSTER v4.1.0 software[82].

**References**

1. Liu, A. Y. & Cohen, M. L. Prediction of New Low Compressibility Solids. *Science* **245**, 841–842 (1989).

2. Occelli, F., Loubeyre, P. & LeToullec, R. Properties of diamond under hydrostatic pressures up to 140 GPa. *Nat. Mater.* **2**, 151–154 (2003).

3. Hart, J. N., Claeyssens, F., Allan, N. L. & May, P. W. Carbon nitride: Ab initio investigation of carbon-rich phases. *Phys. Rev. B* **80**, 174111 (2009).

4. Stavrou, E. *et al.* Synthesis of ultra-incompressible $sp^3$-hybridized carbon nitride with 1:1 stoichiometry. *Chem. Mater.* **28**, 6925–6933 (2016).

5. Muhl, S. & Mendez, J. M. A review of the preparation of carbon nitride films. *Diam. Relat. Mater.* **8**, 1809–1830 (1999).

6. Zinin, P. V. *et al.* Synthesis of new cubic $C_3N_4$ and diamond-like $BC_3$ phases under high pressure and high temperature. *J. Phys. Conf. Ser.* **121**, 062002 (2008).

7. Ming, L. C. *et al.* A cubic phase of $C_3N_4$ synthesized in the diamond-anvil cell. *J. Appl. Phys.* **99**, 033520 (2006).

8. Wang, Y., Liu, Q. & Wang, W. P. Recover of $C_3N_4$ nanoparticles under high-pressure by shock wave loading. *Ceram. Int.* **44**, 19290–19294 (2018).

9. Churkin, V., Kulnitskiy, B., Zinin, P., Blank, V. & Popov, M. The Effect of Shear Deformation on C-N Structure under Pressure up to 80 GPa. *Nanomaterials* **11**, 828 (2021).

10. Goglio, G., Foy, D. & Demazeau, G. State of Art and recent trends in bulk carbon nitrides synthesis. *Mater. Sci. Eng. R* **58**, 195–227 (2008).

11. Fang, L., Ohfuji, H., Shinmei, T. & Irifune, T. Experimental study on the stability of graphitic $C_3N_4$ under high pressure and high temperature. *Diam. Relat. Mater.* **20**, 819–825 (2011).

12. Li, Z. *et al.* Pentadiamond-like Metallic Hard Carbon Nitride. *J. Phys. Chem. C* **124**, 24978−24983 (2020).

13. Wu, Q. *et al.* Ground-state structures, physical properties and phase diagram of carbon-rich nitride $C_5N$. *J. Phys. Condens. Matter* **30**, 385402 (2018).

14. Gao, X., Yin, H., Chen, P. & Liu, J. Shock-induced phase transition of g-$C_3N_4$ to a new $C_3N_4$ phase. *J. Appl. Phys.* **126**, 155901 (2019).

15. Dong, H., Oganov, A. R., Zhu, Q. & Qian, G.-R. The phase diagram and hardness of carbon nitrides. *Sci. Rep.* **5**, 9870 (2015).

16. Li, Z. *et al.* Superhard carbon-rich C-N compounds hidden in compression of the mixture of carbon black and tetracyanoethylene. *Carbon N. Y.* **184**, 846–854 (2021).

17. Wu, Q. *et al.* Unexpected ground-state structures and properties of carbon nitride $C_3N$ at ambient and high pressures. *Mater. Des.* **140**, 45–53 (2018).

18. Pickard, C. J., Salamat, A., Bojdys, M. J., Needs, R. J. & McMillan, P. F. Carbon nitride frameworks and dense crystalline polymorphs. *Phys. Rev. B* **94**, 094104 (2016).

19. Andreyev, A., Akaishi, M. & Golberg, D. Synthesis of nanocrystalline nitrogen-rich carbon nitride powders at high pressure. *Diam. Relat. Mater.* **11**, 1885–1889 (2002).

20. Nesting, D. C. & Badding, J. V. High-Pressure Synthesis of sp$^2$-Bonded Carbon Nitrides. *Am. Chem. Soc.* **4756**, 1535–1539 (2000).

21. Kojima, Y. & Ohfuji, H. Structure and stability of carbon nitride under high pressure and high temperature up to 125GPa and 3000K. *Diam. Relat. Mater.* **39**, 1–7 (2013).

22. Chen, Z. Y., Zhao, J. P., Yano, T. & Ooie, T. Raman characteristics of carbon nitride synthesized by nitrogen-ion-beam-assisted pulsed laser deposition. *Appl. Phys. A Mater. Sci. Process.* **74**, 213–216 (2002).

23. Teter, D. M. & Hemley, R. J. Low-Compressibility Carbon Nitrides. *Science* **271**, 53–55 (1996).

24. Li, Q. *et al.* A novel low compressible and superhard carbon nitride: Body-centered tetragonal $CN_2$. *Phys. Chem. Chem. Phys.* **14**, 13081 (2012).

25. Laniel, D. *et al.* High pressure study of a highly energetic nitrogen-rich carbon nitride, cyanuric triazide. *J. Chem. Phys.* **141**, 234506 (2014).

26. Kessler, F. K. *et al.* Functional carbon nitride materials-design strategies for electrochemical devices. *Nat. Rev. Mater.* **2**, (2017).

27. May, P. W. The New Diamond Age? *Science* **319**, 1490–1491 (2008).

28. *Power Electronics Device Applications of Diamond Semiconductors*. (Elsevier, 2018). doi:10.1016/C2016-0-03999-2

29. Kroke, E. High-Pressure Syntheses of Novel Binary Nitrogen Compounds of Main Group Elements. *Angew. Chemie Int. Ed.* **41**, 77–82 (2002).

30. Horvath-Bordon, E. *et al.* High-Pressure Synthesis of Crystalline Carbon Nitride Imide, $C_2N_2$(NH). *Angew. Chemie Int. Ed.* **46**, 1476–1480 (2007).

31. Ravindran, T. R. & Badding, J. V. UV Raman studies on carbon nitride structures. *J. Mater. Sci.* **41**, 7145–7149 (2006).

32. Kantor, I. *et al.* BX90: A new diamond anvil cell design for X-ray diffraction and optical measurements. *Rev. Sci. Instrum.* **83**, 125102 (2012).

33. Hahn, H., Frank, G. & Klingler, W. Zur Struktur des beta-$Cu_2HgJ_4$ und des beta-$Ag_2HgJ_4$. *Zeitschrift für Anorg. und Allg. Chemie* **279**, 271–280 (1955).

34. München, O. V., Schneider, J. & Haussühl, S. Structure refinements of β-$Si_3N_4$ at temperatures up to 1360°C by X-ray powder investigation. *Zeitschrift für Krist. - Cryst. Mater.* **209**, 328–333 (1994).

35. Liebau, F. *Structural Chemistry of Silicates*. (Springer Berlin Heidelberg, 1985). doi:10.1007/978-3-642-50076-3

36. Jürgens, B. *et al.* Melem (2,5,8-triamino-tri-s-triazine), an important intermediate during condensation of melamine rings to graphitic carbon nitride: synthesis, structure determination by X-ray powder diffractometry, solid-state NMR, and theoretical studies. *J. Am. Chem. Soc.* **125**, 10288–10300 (2003).

37. Duffin, W. J. & Hogg, J. H. C. Crystalline phases in the system In–$In_2S_3$. *Acta Crystallogr.* **20**, 566–569 (1966).

38. Eremets, M. I., Gavriliuk, A. G., Trojan, I. A., Dzivenko, D. A. & Boehler, R. Single-bonded

cubic form of nitrogen. *Nat. Mater.* **3**, 558–563 (2004).

39. Laniel, D. *et al.* High-pressure polymeric nitrogen allotrope with the black phosphorus structure. *Phys. Rev. Lett.* **124**, 216001 (2020).

40. Fujihisa, H. *et al.* Incommensurate Structure of Phosphorus Phase IV. *Phys. Rev. Lett.* **98**, 175501 (2007).

41. Laniel, D. *et al.* Revealing Phosphorus Nitrides up to the Megabar Regime: Synthesis of alpha-$P_3N_5$, delta-$P_3N_5$ and $PN_2$. *arXiv* **2208.08814**, 1–16 (2022).

42. Laniel, D. Cubic Gauche Nitrogen Experimental Crystal Structure Determination. *CCDC 1946895* (2019). doi:10.5517/ccdc.csd.cc23bx2g

43. Wang, X. Polymorphic phases of sp3-hybridized superhard CN. *J. Chem. Phys.* **137**, 184506 (2012).

44. Angel, R. J., Alvaro, M., Gonzalez-Platas, J. & Alvaro, M. EosFit7c and a Fortran module (library) for equation of state calculations. *Zeitschrift fur Krist.* **229**, 405–419 (2014).

45. Datchi, F., Dewaele, A., Le Godec, Y. & Loubeyre, P. Equation of state of cubic boron nitride at high pressures and temperatures. *Phys. Rev. B* **75**, 1–9 (2007).

46. Cohen, M. L. Calculation of bulk moduli of diamond and zinc-blende solids. *Phys. Rev. B* **32**, 7988–7991 (1985).

47. Lyakhov, A. O. & Oganov, A. R. Evolutionary search for superhard materials: Methodology and applications to forms of carbon and $TiO_2$. *Phys. Rev. B* **84**, 2–5 (2011).

48. Chen, X. Q., Niu, H., Li, D. & Li, Y. Modeling hardness of polycrystalline materials and bulk metallic glasses. *Intermetallics* **19**, 1275–1281 (2011).

49. Oganov, A. R. & Glass, C. W. Crystal structure prediction using ab initio evolutionary techniques: Principles and applications. *J. Chem. Phys.* **124**, 244704 (2006).

50. Slack, G. A. Nonmetallic crystals with high thermal conductivity. *J. Phys. Chem. Solids* **34**, 321–335 (1973).

51. Clark, C. D., Dean, P. J. & Harris, P. V. Intrinsic edge absorption in diamond. *Proc. R. Soc. London. Ser. A. Math. Phys. Sci.* **277**, 312–329 (1964).

52. MacDonald, A. H. Bilayer Graphene's Wicked, Twisted Road. *Physics (College. Park. Md).* **12**, 12 (2019).

53. Laniel, D., Weck, G., Gaiffe, G., Garbarino, G. & Loubeyre, P. High-Pressure Synthesized Lithium Pentazolate Compound Metastable under Ambient Conditions. *J. Phys. Chem. Lett.* **9**, 1600–1604 (2018).

54. Xu, Y. *et al.* $LiN_5$: A novel pentazolate salt with high nitrogen content. *Chem. Eng. J.* **429**, 132399 (2022).

55. Zhuang, H. *et al.* Synthesis and Stabilization of Cubic Gauche Polynitrogen under Radio-Frequency Plasma. *Chem. Mater.* (2022). doi:10.1021/acs.chemmater.2c00689

56. Akahama, Y. & Kawamura, H. Pressure calibration of diamond anvil Raman gauge to 410 GPa. *J. Phys. Conf. Ser.* **215**, 012195 (2010).

57. Anzellini, S., Dewaele, A., Occelli, F., Loubeyre, P. & Mezouar, M. Equation of state of rhenium

and application for ultra high pressure calibration. *J. Appl. Phys.* **115**, 043511 (2014).

58. Fedotenko, T. *et al.* Laser heating setup for diamond anvil cells for in situ synchrotron and in house high and ultra-high pressure studies. *Rev. Sci. Instrum.* **90**, 104501 (2019).

59. Frost, D. *et al.* A new large-volume multianvil system. *Phys. Earth Planet. Inter.* **143–144**, 507–514 (2004).

60. Rigaku Oxford Diffraction. CrysAlisPro Software system. (2015).

61. Aslandukov, A., Aslandukov, M., Dubrovinskaia, N. & Dubrovinsky, L. Domain Auto Finder (DAFi) program: the analysis of single-crystal X-ray diffraction data from polycrystalline samples. *J. Appl. Crystallogr.* (2022)--Accepted Manuscript.

62. Dolomanov, O. V., Bourhis, L. J., Gildea, R. J., Howard, J. A. K. & Puschmann, H. OLEX2: A complete structure solution, refinement and analysis program. *J. Appl. Crystallogr.* **42**, 339–341 (2009).

63. Petrícek, V., Dušek, M. & Palatinus, L. Crystallographic computing system JANA2006: General features. *Zeitschrift fur Krist.* **229**, 345–352 (2014).

64. Bykova, E. Single-crystal X-ray diffraction at extreme conditions in mineral physics and material sciences. (University of Bayreuth, 2015).

65. Spahr, D. *et al.* Tetrahedrally Coordinated $sp^3$-Hybridized Carbon in $Sr_2CO_4$ Orthocarbonate at Ambient Conditions. *Inorg. Chem.* **60**, 5419–5422 (2021).

66. Zurkowski, C. C., Lavina, B., Chariton, S., Prakapenka, V. & Campbell, A. J. Stability of $Fe_2S$ and $Fe_{12}S_7$ to 125 GPa; implications for S-rich planetary cores. *Geochemical Perspect. Lett.* **21**, 47–52 (2022).

67. Zhang, L., Yuan, H., Meng, Y. & Mao, H. K. Development of High-Pressure Multigrain X-Ray Diffraction for Exploring the Earth's Interior. *Engineering* **5**, 441–447 (2019).

68. Holtgrewe, N., Greenberg, E., Prescher, C., Prakapenka, V. B. & Goncharov, A. F. Advanced integrated optical spectroscopy system for diamond anvil cell studies at GSECARS. *High Press. Res.* **39**, 457–470 (2019).

69. Giannozzi, P. *et al.* QUANTUM ESPRESSO: a modular and open-source software project for quantum simulations of materials. *J. Phys. Condens. Matter* **21**, 395502 (2009).

70. Giannozzi, P. *et al.* Advanced capabilities for materials modelling with Quantum ESPRESSO. *J. Phys. Condens. Matter* **29**, 465901 (2017).

71. Giannozzi, P. *et al.* Quantum ESPRESSO toward the exascale. *J. Chem. Phys.* **152**, 154105 (2020).

72. Blöchl, P. E. Projector augmented-wave method. *Phys. Rev. B* **50**, 17953–17979 (1994).

73. Perdew, J. P., Burke, K. & Ernzerhof, M. Generalized Gradient Approximation Made Simple. *Phys. Rev. Lett.* **77**, 3865–3868 (1996).

74. Grimme, S., Antony, J., Ehrlich, S. & Krieg, H. A consistent and accurate ab initio parametrization of density functional dispersion correction (DFT-D) for the 94 elements H-Pu. *J. Chem. Phys.* **132**, 154104 (2010).

75. Monkhorst, H. J. & Pack, J. D. Special points for Brillouin-zone integrations. *Phys. Rev. B* **13**,

5188–5192 (1976).

76. Togo, A. & Tanaka, I. First principles phonon calculations in materials science. *Scr. Mater.* **108**, 1–5 (2015).

77. Hellman, O., Abrikosov, I. A. & Simak, S. I. Lattice dynamics of anharmonic solids from first principles. *Phys. Rev. B* **84**, 2–5 (2011).

78. Hellman, O. & Abrikosov, I. A. Temperature-dependent effective third-order interatomic force constants from first principles. *Phys. Rev. B - Condens. Matter Mater. Phys.* **88**, 1–5 (2013).

79. Oganov, A. R. & Glass, C. W. Crystal structure prediction using ab initio evolutionary techniques: Principles and applications. *J. Chem. Phys.* **124**, 244704 (2006).

80. Heyd, J., Scuseria, G. E. & Ernzerhof, M. Hybrid functionals based on a screened Coulomb potential. *J. Chem. Phys.* **118**, 8207–8215 (2003).

81. Müller, P. C., Ertural, C., Hempelmann, J. & Dronskowski, R. Crystal Orbital Bond Index: Covalent Bond Orders in Solids. *J. Phys. Chem. C* **125**, 7959–7970 (2021).

82. Nelson, R. *et al.* LOBSTER: Local orbital projections, atomic charges, and chemical-bonding analysis from projector-augmented-wave-based density-functional theory. *J. Comput. Chem.* **41**, 1931–1940 (2020).

**Data availability**

Structural data was deposited at the Cambridge Crystallographic Data Centre (CCDC), CSD 2202353-2202360. All other datasets generated during and/or analysed during the current study are available from the corresponding author on reasonable request.

**Acknowledgments**

The authors acknowledge the European Synchrotron Radiation Facility (ESRF) for the provision of beamtime at the ID27 and ID11 beamlines, as well as the Advanced Photon Source for beamtime at the GSECARS beamline. D.L. thanks the Deutsche Forschungsgemeinschaft (DFG, project LA-4916/1-1) and the UKRI Future Leaders Fellowship (MR/V025724/1) for financial support. N.D. and L.D. thank the Federal Ministry of Education and Research, Germany (BMBF, grant no. 05K19WC1) and the Deutsche Forschungsgemeinschaft (DFG; projects DU 954–11/1, DU 393–9/2, and DU 393-13/1) for financial support. B.W. gratefully acknowledges funding by the DFG in the framework of the research unit DFG FOR2125 and within projects WI1232 and thanks BIOVIA for support through the Science Ambassador program. N.D., I.A.A. and Fe.T. also thank the Swedish Government Strategic Research Area in Materials Science on Functional Materials at Linköping University (Faculty Grant SFO-Mat-LiU No. 2009 00971). I.A.A. and Fl.T. are supported by the Swedish Research Council (VR) Grant No. 2019-05600. I.A.A. acknowledges support from the Knut and Alice Wallenberg Foundation (Wallenberg Scholar grant no. KAW-2018.0194). W.S. acknowledges funding support from the Deutsche Forschungsgemeinschaft (DFG, German Research Foundation) under Germany's Excellence Strategy-EXC 2089/1-390776260 (e-conversion). A.V.P. acknowledges support from the RSF grant № 22-12-00193. Computations were enabled by resources provided by the Swedish National Infrastructure for Computing (SNIC) using Dardel at the PDC Center for High Performance Computing, KTH Royal Institute of Technology and LUMI at the IT Center for Science (CSC), Finland through grant SNIC 2022/6-10 and SNIC 2021/37-10, respectively. For the purpose of open access, the author has applied a Creative Commons Attribution (CC BY) licence to any Author Accepted Manuscript version arising from this submission.

**Author Contributions**

D.L., L.D. and N.D. designed the work. D.L. and L.D. prepared the high-pressure experiments. A.N. prepared the black phosphorus precursor. D.L., A.A., S.K., T.F., Y.Y., F.I.A., S.C., C.G., J.W., G.G., B.W., A.P., M.M. and V.P. performed the synchrotron X-ray diffraction experiments. D.L. and L.D. processed the synchrotron X-ray diffraction data. D.L., L.D. and G.W. performed the Raman spectroscopy measurements and analyzed the data. Fl.T., Fe.T., A.V.P, B.W., V.M. and I.A.A. performed the theoretical calculations. D.L., Fl.T., L.D., N.D., I.A.A. and W.S. prepared the first draft of the manuscript with contributions from all other authors. All the authors commented on successive drafts and have given approval to the final version of the manuscript.

**Competing Interests**

The authors declare no competing interests.

**Materials and Correspondence**

Dominique Laniel (dominique.laniel@ed.ac.uk) should be contacted for additional requests.